# A Brief History of Context

Kaiyu Wan

**Computer Science Department, East China Normal University**
**Shanghai, 200, China**

## Abstract

Context is a rich concept and is an elusive concept to define. The concept of context has been studied by philosophers, linguists, psychologists, and recently by computer scientists. Within each research community the term context was interpreted in a certain way that is well-suited for their goals, however no attempt was made to define context. In many areas of research in computer science, notably on web-based services, human-computer interaction (HCI), ubiquitous computing applications, and context-aware systems there is a need to provide a formal operational definition of context. In this brief survey an account of the early work on context, as well as the recent work on many working definitions of context, context modeling, and a formalization of context are given. An attempt is made to unify the different context models within the formalization. A brief commentary on the usefulness of the formalization in the development of context-aware and dependable systems is included.

## 1. Introduction

According to the Oxford English Dictionary (OED), context denotes "the circumstances that form the setting for an event". To emphasize a common social usage of the word context OED includes the quotation [12] "I wish honorable gentlemen would have the fairness of what I did say, and not pick out detached words". Although the word context has been used for a long time in many scientific descriptions, literary essays, and in philosophical discourses, its meaning was always left to the reader's understanding. In one of the earlier papers, Clark and Carlson [11] state that Context has become a favorite word in the vocabulary of cognitive psychologists and that it has appeared in the titles of a vast number of articles. They then complain that the denotation of the word has become murkier as its uses have been extended in many directions and deliver the now widespread opinion that context has become some sort of "conceptual garbage can". That context has changed now. The importance of context in information retrieval, knowledge representation, reasoning in AI, and analysis of computer programs have been recognized and there is a serious effort to make a precise technical working definition of the notion of context. More recently, the importance of context was picked up by researchers in many areas of computer science, most importantly those working in Human-Computer Interaction (HCI), semantic web, and trustworthy systems. This intense interest has produced many operational definitions of context, but almost all of them are either informal or use ad hoc notation. We review in this paper the different types of notations and interpretations used for context. The review is classified into *Context in Logic*, *Context in Languages*, and *Context in Systems*. This classification and review are not exhaustive. It is used mainly to trace the historical progression of the systematic study of context in different, but related, areas.

### 1.1 Structure and Interpretation

The word "context" is derived from the Latin words *con* (meaning "together") and *texere* (meaning "to weave"). The raw meaning of it is therefore "weaving together". A circumstance is a weaving together of many types of entities. Thus, in describing a context we must define a finite set of entities, a finite set of properties for each entity, and the inter-weaving of the properties. As an example, the setting for a "seminar event" is the weaving together of the entities *speaker*, *topic*, *audience*, *time*, *location* and their properties such as *name* and *affiliation* for the speaker, *title* and *abstract* for the topic, *size* and *status* for audience, *clocktime* for time, and *building-address* and *room-number* for locality. We need to associate with each property a value from its domain and bind each entity with the instantiated properties in order to describe the context of seminar. The choice of entities, the choice of properties, and the notation used for binding them are crucial for system development. This choice for context definition has the effect of narrowing down the possible interpretations of declared policies and constraints for system development. Context description also eliminates ambiguities. It should be possible to define contexts in programming languages independent of how it should be used. For example, if a context is defined by *locality* and *time*, then many events may happen in a specific context, and each event may produce a different experience in one context. Therefore, the structural definition of a context is only part of its specification, its other part being the semantics of the *world* associated with the context. The world may be defined by a set of states in programming or by a set of logical formulas in a formal specification. Motivated by a need to specify context as a first class citizen in languages and systems, a formal representation of it was developed in [45].

## 2. Context in Logic

In this section we review the study of context in logic as a formal object and reasoning. We review intensional logic and some variations of propositional and predicate logic in which context has been embedded as first class citizens.

### 2.1 Intensional Logic

Intensional Logic [14, 42], a family of mathematical formal systems that permits expressions whose value depends on *hidden context*, came into being from research in natural language understanding. According to Carnap [9], the real meaning of a natural language expression whose truth-value depends on the context in which it is uttered is its *intension*. The *extension* of that expression is its actual truth-value in the different possible contexts of utterance, where this expression can be evaluated. Basically, intensional logics add *dimensions* to logical expressions, and non-intensional logics can be viewed as *constant* in all possible dimensions, i.e. their valuation does not vary according to their context of utterance. *Intensional operators* are defined to *navigate* in the context space. In order to navigate, some dimension *tags* (or indexes) are required to provide placeholders along dimensions. These dimension tags, along with the dimension names they belong to, are used to define the context for evaluating intensional expressions.

**Example 1**  E: Beijing is now the capital of China.
This expression is intensional because the truth value of this expression depends on the context in which it is evaluated. The intensional natural language operator in this expression is now, which refers to the time dimension. Today it is certainly true, but there existed time points in the past when China had a different capital. For example, before 1949, the capital of China was NanJing. Those different values (i.e. True or False) along different time points are extensions of this expression. In other words, the evaluation of the above expression is time-dependent. A natural extension is to consider expressions that depend on more than one dimension, such as time, space, audience, and so on.

**Example 2**  The meaning of the expression:
E: the overseas indexes during this period close 10% below their highs can be interpreted when the possible worlds spanned by the dimensions overseas and period are defined. The Table below gives a possible extension of the expression E when the periods are months in a given year, and overseas stock markets are Amsterdam, Brussels, Frankfurt, and London. By varying the year we get 3-dimensional extension.

Table 1: Example 1

|          | Ja | Fe | Mr | Ap | Ma | Jn | Jl | Au | Se | Oc | No | De |
|----------|----|----|----|----|----|----|----|----|----|----|----|----|
| Amsterda | F  | F  | F  | F  | T  | T  | T  | T  | T  | F  | F  | F  |
| Brussels | F  | F  | F  | T  | T  | T  | T  | T  | T  | F  | F  | F  |
| Frankfurt| F  | F  | T  | T  | T  | T  | T  | T  | T  | T  | F  | F  |
| London   | F  | T  | T  | T  | T  | T  | T  | T  | T  | T  | T  | T  |

### 2.2 Formalizing Context in AI

Contexts in AI were introduced by Weyhrauch (1980) [51] and subsequently developed by McCarthy and Buva c (1998) [28] and Giunchiglia (1993) [17]. Surveys of the formalizations and the usage of contexts can be found in Sharma (1995) [38], Akman and Surav (1996) [2], Bouquet et al. (1999) [5], Bonzon et al. (2000) [6] and Akman et al. (2001) [3]. Context serves an important purpose in AI and Intelligent Information Processing (IIP). The classic example of the earliest IIP that failed to meet safety criteria is MYCIN [39]. It was observed by McCarthy [2]. MYCIN system advises physicians on treating bacterial infections of the blood and meningitis. When MYCIN was first introduced context was not part of system's query processing phase. When it was given the query "what is the treatment for *Chlorae Vibrio*" it recommended "two weeks of *tetracycline*" treatment. What it failed to inform the physician was that a massive *dehydration* during the course of the treatment would occur. While the administration of *tetracycline* would cure the bacteria, the patient would die long before that due to diarrhea. Here is an instance where the context of correct usage was not given to the physician, which ultimately made the system *unsafe* and hence not trustworthy. A contextual MYCIN will explicitly state the context for correct administration of medications. In AI context is formalized using propositional or predicate logic.

Contexts are abstract objects (representation free) and are first-class citizens. Consequently, contexts are freely used in logical formulas, without explicitly defining contexts. In some sense, in the logical approach a context itself is defined by a set of formulas that are true by the truth assignments in that context. In [28] McCarthy, who introduced a logical framework in AI for studying context, gave three reasons justifying his approach.

- *Axiomatization:* The use of contexts simplifies axiomatizations. Axioms from one context can be *lifted* to more general contexts.
- *Vocabulary and Interpretation:* Contexts allow the use of specific vocabulary and information. Terms that are used in one context have particular meaning, which they will not have in general.
- *Building AI Systems:* A hierarchy of AI systems can be built be transcending from one context to another.

According to Giunchiglia (1993) [17], the notion of context formalizes the idea of localization of knowledge and reasoning. Intuitively speaking, a context is a set of facts (expressed in a suitable language, usually different for each different set of facts) used locally to prove a given goal, plus the inference routines used to reason about them (which can be different for different sets of facts). A context encodes a perspective about the world. It is a partial perspective as the complete description of the world is given by the set of all the contexts. It is an approximate perspective, in the sense described in McCarthy (1979) [27], as we never describe the world in full detail. Finally, different contexts, in general, are not independent of one another as the different perspectives are about the same world, and, as a consequence, the facts in a context are related to the facts in other contexts.

The work in Giunchiglia and Serafini (1994) provides a logic, called Multi Language Systems (ML Systems), formalizing the principles of reasoning with contexts informally described in Giunchiglia (1993). In ML systems, contexts are formalized using multiple distinct languages, each language being associated with its own theory (a set of formulas closed under a set of inference rules). Relations among different contexts are formalized using bridge rules, namely inference rules with premises and consequences in distinct languages. Recently, Ghidini and Giunchiglia (2001) proposed Local Models Semantics (LMS) as a model-theoretic framework for contextual reasoning, and use ML systems to axiomatize many important classes of LMS. From a conceptual point of view, Ghidini and Giunchiglia argued that contextual reasoning can be analyzed as the result of the interaction of two very general principles: the principle of locality (reasoning always happens in a context); and the principle of compatibility (there can be relationships between reasoning processes in different contexts). In other words, contextual reasoning is the result of the (constrained) interaction between distinct local structures. A good survey of context formalization in AI and a comparison between different formalizations can be found in [2]. According to this exposition, context is either treated within some logical framework or within situation theory. Both approaches deal with abstract contexts and focus only on contextual reasoning.

# 3. Context in Languages

We review the role of context in intensional programming languages (IPL) and in $\lambda$ calculus.

### 3.1 Formalizing Context in AI

The intensional programming paradigm has its foundations on intensional logic. It retains two aspects from intensional logic: first, at the syntactic level, are contextswitching operators, called *intensional operators*; second, at the semantic level, is the use of *possible world semantics*. By making difference between intension and extension, IPL provides two different levels for programming. On the higher level, it allows to represent/express problems in a declarative manner; on the lower level, it solves problems without loss of accuracy.

IPL deals with *streams* of entities which could be numbers, or strings of characters, or any computable structure. These streams are first class objects in intensional languages and functions can be applied to these streams. Because of the infinite nature of IPL, it is especially appropriate for describing the behavior of systems that change with time or physical phenomena that depend on more than one parameter (such as time, space, temperature, etc). It is also an appropriate language for use in business applications that generate data streams, or textual streams, or media streams. Examples include stock market transactions and credit card transactions which are mostly data streams of records where each record contains information on a transaction, call center transactions that generate textual streams of conversations, and multi-media streams that are generated by cable companies to distribute movies on demand. The streams are processed by accessing certain semantic units and interpreting it in different contexts.

There is no notion of type in an IPL. The operators on the stream contents are assumed to be given when one writes the stream functions. The natural logical view of a stream is an infinite sequence, and in writing programs one does not worry about the physical representations of stream contents. This abstraction enables one to understand an IPL program from the statements in it, without any reference to its implementation. The computational model for IPLs is known as *eduction*. That is, an implementation starts computing the first element that satisfies a given context, then the second, and so on. A context for expression evaluation, as informally understood in Example 2, is described by a set of dimensions (attributes) and a *finality* (goal). The finality is domain-dependent and is chosen so that a finite set of dimensions would suffice to realize that goal. For example in processing call center streams *understanding a conversation* may be the finality, and the attributes may be a set of *key words* chosen in advance to meet the goal. As another example, in processing streams of user interactions with web, the finality may be *understanding user patterns* and the attributes may be *ActivityLocation*, *ActivityDuration*, and *VolumeofDataTransfer*. Both finality and the attributes defining a context are implicitly used in evaluating the extensions from a stream.

## 3.2 Lambda Calculus

In programming languages, context is a meta concept: *static* context introduces constants, definitions, and constraints, and *dynamic* context processes the executable information for evaluating expressions. In [35] context is introduced in the lambda calculus and an argument is made for introducing context as first class objects in programming languages. Their motivation for introducing context in the theory of lambda calculus is to develop a programming language with first-class contexts that has advanced programming features for manipulating open terms. We are motivated along similar lines for introducing context in Lucid. However there are significant differences in the semantics of context between the two approaches.

A context in the lambda calculus is defined as a term with a "hole" in it. The hole in a context can be filled with a term which may involve free variables. To avoid inconsistent hole filling within the scope of lambda binding the holes are labeled, hole abstraction, and context application are separated. Informally, a context C with a hole in it, written C[●], will become the term C[M] when the hole is filled with the term M. The formal way of writing this in $^\lambda$ calculus is M'$\circ$M , where the term M' abstracts the hole in the context. The term M that abstracts the hole labeled X itself is written as $\delta$X.M'. For example, the context C[●] = ($^\lambda$ x.[●] +y)3 is represented by the term M'= ($\delta$X($^\lambda$ x.X + y)3). The term obtained by filling the hole in M' with x+ z is written ($\delta$X.($^\lambda$ x.X+y)3) $\circ$ (x+z).

In our theory context plays two roles: one role is as a reference to an item in a multi-dimensional stream, and the other role is as a descriptor of situations at which expressions are evaluated. A stream of contexts may be constructed and context expression may be evaluated at a context. In Lucx expressions and contexts exist independently. A context may be defined without any regard to any specific expression and hence it may be used to evaluate different Lucx expressions. Similarly, an expression can be evaluated at different contexts. It is possible to define context dependent expressions in Lucx. Such expressions may be evaluated at a context distinct from any other context used in its definition. We can define nested contexts, and dependent contexts. These features offer a variety of flexible ways to programming different applications.

## 3.3 Lucid and Lucx

Lucid was originally invented as a *Program Verification Language* by Ashcroft and Wadge [1]. And later it evolved into a dataflow language [52]. The basic intensional operators are *first*, *next*, and *fby*. The four operators derived from the basic ones are *wvr*, *asa*, *upon*, and *prev*, where *wvr* stands for *whenever*, *asa* stands for *as soon as*, *upon* stands for *advances upon*, and *prev* stands for *previous*. Lucid is a *stream* (i.e. infinite entity) manipulation language. All the above operators are applied to streams to produce new streams. The definitions of these operators [30] are shown as follows

**Definition 1** If X = (x0, x1,…, xi,… ) and Y = (y0, y1,…, yi,…), then

(1) $\underline{first}$ X = (x0, x0,…, x0,…)

(2) $\underline{next}$ X    = (x1, x2,…, xi+1,…)

(3) X $\underline{fby}$ Y = (x0, y0, y1,…, yi-1,…)

(4) X $\underline{wvr}$ Y = if $\underline{first}$ Y then X
                $\underline{Fby}$ (nextX $\underline{wvr}$ $\underline{next}$ Y)
              Else (next X $\underline{wvr}$ $\underline{next}$ Y)

(5) X $\underline{asa}$ Y = $\underline{first}$ (X $\underline{wvr}$ Y)

(6) X $\underline{upon}$ Y = X $\underline{fby}$ (if $\underline{first}$ Y
             then (next X upon next Y)
             else (X upon next Y))

(7) $\underline{prev}$ X     = X@(#1) 2

Example 3 illustrates the definitions on a stream *A* whose elements are integers, and a stream *B* whose elements are boolean. In a boolean stream the symbols 1 and 0 indicate true and false respectively. The symbol *nil* indicates an undefined value.

**Example 3** :

| | | | | | | |
|---|---|---|---|---|---|---|
| *A* | = | 1 | 2 | 3 | 4 | 5 |
| *B* | = | 0 | 0 | 1 | 0 | 1 |
| *first A* | = | 1 | 1 | 1 | 1 | 1 |
| *next A* | = | 2 | 3 | 4 | 5 | |
| *prev A* | = | *nil* | 1 | 2 | 3 | 4 | 5 |
| *AfbyB* | = | 1 | 0 | 0 | 1 | 0 | 1 |
| *A wvr B* | = | 3 | 5 | | | |
| *A asa B* | = | 3 | 3 | 3 | | |
| *A upon B* | | 1 | 1 | 1 | 3 | 3 | 5 |

With the operators defined above, Lucid only allows sequential access into streams. That is, the $(i + 1)th$ element in a stream is only computed once the $ith$ element has been computed. To enable subcomputations to take place in arbitrary dimensions and all indexical operators to be parameterized by one or several dimensions, two basic intensional operators are added. One is *intensional navigation* (*@.d*), which allows the values of a stream to vary along the dimension *d*. Another is *intensional query* (*#.d*), which refers to the current position (i.e. tag value) along the dimension *d*. This way, it is possible to access streams randomly.

Example 4 illustrates the definitions of these two operators on two streams *A* and *B* along the *time* dimension.

Example 4

$A = 1\ 2\ 4\ 8\ 16\ 32\ 64\ 128\ …$
$B = 1\ 2\ 3\ 0\ 6\ 7\ 4\ 5\ …$
$A\ @.time\ B = 2\ 4\ 8\ 1\ 64\ 128\ 16\ 32\ ...$
$#.time = 0\ 1\ 2\ 3\ 4\ 5\ 6\ 7\ ...$

The major distinction between contexts in AI and in IPL is that in the former case they are *rich objects* that are not *completely expressible* and in the later case they are *implicitly* expressible. Hence it is possible to write an expression in Lucid whose evaluation is context-dependent. However, a context in the current version of Lucid can not be explicitly manipulated. This restricts the ability of Lucid to be an effective programming language for programming diverse applications. So we have extended Lucid by adding the capability to explicitly manipulate contexts. This is achieved by introducing *context* as a first class object in the language. That is, contexts can be declared, assigned values, used in expressions, and passed as function parameters. The language thus extended, is called as *Lucx* [45] (Lucid extended with *contexts*)(the x is used as the x in TeX). Thus, the rationale for introducing context in Lucid is quite analogous to the introduction of context to enrich knowledge base in AI. However, our notion of context differs significantly from McCarthy's. In our study context is both *finite* and *concrete*. It is finite in the sense that only a finite number of dimensions are allowed in defining a context. However it does not impose any limitation on handling infinite streams, because with every dimension an infinite *tag set* is introduced in the language. A full account of context-based evaluation of expressions in Lucx is given in [45].

## 4. Context in Systems

Context-aware adaptation is regarded as the most important feature for pervasive and ubiquitous services [50, 20, 25]. Web services [26] and mobile computing applications [53, 24] immensely benefit with a formal context model. It is in this context that we review the role of context. Context modeling and context-dependent interpretive actions are important in HCI [13, 15, 49]. However in all these works context is not formalized. In this section, after we review context formalism, we explain how our formal definition provides a rigorous platform for developing context-aware systems.

### 4.1 Formalizing Context

We formalize context as *a typed relation*, a set of ordered pairs of $(d, x)$ where $d$ is a dimension, $T_d$ is the type of $d$ and $x : T_d$.

**Definition 2** Let DIM denote the set of all possible dimensions, and T = $\{T_d \mid d \in DIM\}$ be the set of types associated with the dimensions. A context c is a finite relation $\{f(d, x) \mid d \in DIM \wedge x : T_d\}$. The degree of the context c is $|dom\ c|$. The empty relation corresponds to Null context. The degree of Null context is 0.

A context having only one (dimension, tag) pair is called a micro context. Let G denote the set of contexts over {DIM, T}. The set of micro contexts is M = $\{c| c \in G; |c| = 1\}$. The set of simple contexts is S = $\{c| c \in G, c$ is a partial function}. Clearly, a simple context c of degree 1 is a micro context. A context which is not simple is referred to a non-simple context. The basic functions dim and tag are to extract the set of dimensions and the values associated with the dimensions in a context. That is, if c = $f\langle d_1,x_1\rangle,....,\langle d_k,\ x_k\rangle\}$, then we may write c = $\{m_i \mid m_i = \langle d_i,x_i\rangle\}$, dim(c) = $\{d_1, d_2,…d_k\}$, and tag(c) = $\{x_1, x_2,… x_k\}$. For the tuple (d, x) in a micro context c we use the functions $dim_m$ and $tag_m$ to extract the tuple components: $dim_m(c) = d$ and $tag_m(c) = x$.

### 4.2 Context Operators

In this section, context operators are discussed. A context being a relation we borrow the notation and meaning of those relational operators that are available in mathematics. Rest of them we define, using set theory notation. Using these context operators contexts can be managed dynamically and flexibly. The syntax of context expressions are also formally defined. In order to evaluate context expression correctly, precedence rules for context operators are provided as well.

Context operators are : *override* $\oplus$ , *difference* $\ominus$, *choice* / , *conjunction* $\cap$ , *disjunction* $\cup$ , *undirected range* $\rightleftharpoons$ , *directed range* $\rightarrow$ , *projection* $\downarrow$ , *hiding* $\uparrow$ , *substitution* / , *comparison* =, $\subseteq$, $\supseteq$ . The *difference* $\ominus$, *conjunction* $\cap$, *disjunction* $\cup$ , and *comparison* =, $\subseteq$, $\supseteq$ operators are set operators. The rest of the operators are explained and formally defined below.

**Definition 3** <u>Override</u> $\oplus$ This operator takes two contexts $c1 \in$ G, and $c2 \in$ S and returns a context c $\in$ G, which is the result of the conflict -free union of c1 and c2, as defined below:

$$\_ \oplus \_ : G \times S \rightarrow G,$$

c= c1 $\oplus$ c2 ={m | ( m $\in$ c1 $\wedge$ dim$_m$(m) $\notin$ dim(c2)) $\vee$ m $\in$ c2}

**Definition 4** <u>Choice</u> | This operator accepts a finite number of c1ck of contexts and non-deterministically return one of the cis. The definition c= c1|c2|…,|ck implies that c is one of the ci, where $1 \leq i \leq k$:

$$\_|\_ : G \times G \times \ldots \times G \rightarrow G,$$

**Definition 5** <u>Projection.</u> This operator takes a context $c \in$ G and a set of dimensions $D \subset$ DIM as arguments and filters only those micro contexts in c that have their dimensions in set D.

$$\_ \downarrow \_ : G \times D \rightarrow G$$

c $\downarrow$ D ={m| m $\in$ c $\wedge$ dim$_m$(m) $\in$ D}.

**Example 5** :

Let c1 = { ( d,1) , (e,4) , (f,3)}, D={d,e}

then c1 $\downarrow$ D= { ( d,1) , (e,4)}

**Definition 6** <u>Hiding.</u> This operator enables a set of dimensions D to be applied on a context $c \in$ G to remove all the micro context s in c whose dimensions are in D:

$$\_ \uparrow \_ : G \times D \rightarrow G \quad ,$$

c $\uparrow$ D={m| m $\in$ c $\wedge$ dim$_m$(m) $\notin$ D}

**Example 6** :

Let c1 = { ( d,1) , (e,4) , (f,3)}, D={d,e}

then c1 $\uparrow$ D = {(f,3)}

**Definition 7** <u>Substitution.</u> This operator takes a general context and a simple context as arguments and produces a context which is the result of replacing a sub-context of the general context with a sub-context of the simple context if their domains are equal.

$$\_/\_ : G \times S \rightarrow G \quad ,$$

c/s = ( c $\uparrow$ dim s) $\bigcup$ (s $\downarrow$ dim c)

**Example 7** :

Let c1 ={(d, 1), (e, 4),(d, 3)},

c2 ={(d, 4),(f, 3)}, then c1/c2 = {(e,4),(d,4)}

**Definition 8** <u>Undirected</u> range. This operator takes two contexts c1, c2 $\in$ G as arguments and returns a set of simple contexts. The tag set U is assumed to be totally ordered. We give a constructive definition here.

$$\_ = \_ : G \times G \rightarrow P \, S$$

Steps for constructing the final result are shown as follows:

1. Let S' be the set of simple contexts, which is the result of ( c1 = c2).

2. For each pair of m1 $\in$ c1, m2 $\in$ c2, and dim$_m$(m1) = dim$_m$ (m2 ), do the following:

(a) Define a = min{tag$_m$ (m1), tag$_m$ (m2 )} and b = max{tag$_m$(m1), tag$_m$(m2)}

(b) Define the subrange t$_{ba}$ = a..b.

(c) Construct the set Y1:

Y1 = { (d$_{ba}$ ,x)| d$_{ba}$ = dim$_m$(m1) = dim$_m$(m2), x $\in$ t$_{ba}$}

3. Y={Y1,Y2,…Yp}, Where Yi(i = 1,…,p), are the sets of micro context s constructed in Step 2. Define for Yi $\in$ Y, first(Yi) ={dim$_m$(m) | m $\in$ Yi}, and, second(Yi) ={tag$_m$(m)| m $\in$ Yi}. If there exists Yi, Yj$\in$ Y such that first(Yi) = first(Yj), for i$\neq$ j, we replace the sets Yi and Yj by their union Yi $\bigcup$ Yj, and repeat this process until the first(Yi)s for Yi Y are distinct

4. For Yi $\in$ Y, construct the set Z of contexts: Z={{(first(Y1),x1), (first(Y2), x2),…, (first(Yp), xp)}| (x1, x2, …, xp) $\in \prod^p_{i=1}$second(Yi))}.

5. Define: Xc1 = c1 $\uparrow$ $\bigcup_{Yi \in Y}$ first(Yi).

6. Define: Xc2 = c2 $\uparrow$ $\bigcup_{Yi \in Y}$ first(Yi)..

7. Construct S': S'= {{z} $\bigcup$ X$_{c1}$ $\bigcup$ X$_{c2}$ | z $\in$ Z}.

Basically, the result consists of three parts:

1. For each pair m1 $\in$ c1, m2 $\in$ c2 which shares the same dimension, constructs a set Yi, this is done in step 2 and step 3. The result of union the set Yi, done in step 4, consists of the first part : Z.

2. All the other micro context s of c1 which have different dimensions consists of the second part : Xc1.

3. Similarly, all the other micro context s of c2 which have different dimensions consists of the third part : Xc2.

**Example 8** :

*Let c1* ={(*e*, 3), (*d*, 1)}, *c2* ={(*e*, 1),(*d*, 3)},

c3 ={(*e*, 3)},   c4 ={(*f*, 4)},

c5 ={(*e*, 1), (*f*, 4)}

*then c1* = *c2*={{(*e*,1),(*d*,1)}, {(*e*,1),(*d*,2)},

{ (*e*,1),(*d*,3)}, {(*e*,2),(*d*,1)},{(*e*,2),(*d*,2)}, { (*e*,2),(*d*,3)},

{(*e*,3),(*d*,1)}, {(*e*,3),(*d*,2)}, { (*e*,3),(*d*,3)}

c3 = c4 ={(*e*,3),(*f*,4)}

c3= c5 ={{(*e*,1),(*f*,4)},{(*e*,2),(*f*,4)},

{(*e*,3),(*f*,4)}}

**Definition 9** *Directed Range.* This operator takes two contexts c1 $\in$ G and c2 $\in$ S and returns a set of contexts:

$$\_ \overset{\rightarrow}{=} \_ : G \times S \rightarrow P \, G$$

*We change only Step 2 of the method described for the undirected range(Page 7) to obtain the result:*

*(a) Define a = tag$_m$ (m1), b = tag$_m$ (m2 ), if tag$_m$ (m1) < tag$_m$(m2), else ignore the pair m1, m2.*

*(b) Define the subrange t$_{ba}$ = a..b.*

**Example 9** :

*Let c1* = {(*d*,1)}, *c2* = {(*d*,3), (*f*,4)},

*then* $c1 \rightharpoonup c2 = c1 \rightleftharpoons c2 = \{(d,1),(f,4)\}, \{(d,2),(f,4)\}, \{(d,3),(f,4)\}\}$,
$c2 \rightharpoonup c1 = \{(f,4)\}$

## 4.2 Context Expression

Informally, a context expression is an expression involving context variables and context operators. Let $c$ be a context variable, and D be a set of dimensions. A formal syntax for context expression $C$ is shown in Figure 1(left column). A context expression that satisfies those syntactic rules is a well-formed context expression.

In order to provide a precise meaning for a context expression, we define the precedence rules for all the operators. Figure 1(right column) shows the operator precedence from the highest (top row) to the lowest (bottom row). Parentheses will be used to override this precedence when needed. Operators having the same precedence will be applied from left to right.

| syntax | | precedence |
|---|---|---|
| $C$ ::= $c$ | $C = C$ | $\downarrow, \uparrow, /$ |
| \| $C \supseteq C$ | $C \subseteq C$ | \| |
| \| $C \mid C$ | $C/C$ | \| |
| \| $C \oplus C$ | $C \ominus C$ | $\cap, \cup$ |
| \| $C \cap C$ | $C \cup C$ | $\oplus, \ominus$ |
| \| $C \rightleftharpoons C$ | $C \rightarrow C$ | $\rightleftharpoons, \rightarrow$ |
| \| $C \downarrow D$ | $C \uparrow D$ | $=, \subseteq, \supseteq$ |

Fig. 1 Formal Syntax of Context Expressions and Precedence Rules for Context Operators.

**Example 11** *Given a well -formed context expression* $c3 \uparrow D \oplus c1 \mid c2$, *where* $c1 = \{(x,3),(y,4),(z,5)\}$, $c2 = \{(y,5)\}$, *and* $c3 = \{(x,5),(y,6),(w,5)$, $D = \{w\}$, *the evaluation steps are shown as follows:*

[Step1]. $c3 \uparrow D = \{(x,5),(y,6)\}$[Definition 6, Page 6]

[Step2]. $c1 \mid c2 = c1$ *or* $c2$           [Page 6] [Step3].
Suppose in Step2, $c1$ is chosen,

$c3 \uparrow D \oplus c1 = \{(x,3),(y,4),(z,5)\}$ [Definition 3, Page 6]
else if $c2$ is chosen,
$c3 \uparrow D \oplus c2 = \{(x,5),(y,5)\}$[Definition 3, Page 6]

## 4.3 Context Set Operators

In Lucx we avoid higher-order sets of contexts, and allow only sets of simple contexts. Hereafter, by "set of contexts" we refer only to "set of simple contexts". There are two kinds of such operators: *lifting* operators, and *relational* operators.

### 4.3.1 Lifting Operators

**Definition 11** <u>*Projection.*</u> *For* $s \in P\ S$, *and* $D \subseteq DIM$. *The projection operator constructs a set of contexts* $s' \in P\ S$, *where* $s'$ *is obtained by projecting each context from s on to the dimension set D.*

$$\_\downarrow\_ : P\ S\ \times P\ DIM \rightarrow P\ S$$

$s' = s \downarrow D = \{c \downarrow D / c \in s\}$

**Definition 12** <u>*Hiding.*</u> *For* $s \in P\ S$, *and* $D \subseteq DIM$. *The hiding operator constructs* $s' \in S2$, *where* $s'$ *is obtained by hiding each context in s on the dimension set D.*

$$\_\uparrow\_ : P\ S\ \times P\ DIM \rightarrow P\ S$$

$s' = s \uparrow D = \{c \uparrow D / c \in s\}$

**Definition 13** <u>*Substitution.*</u> *This operator produces a set of contexts s, sP S, for a given set of contexts s, s P S, a dimension and a tag value belonging to that dimension:*

$$\_/\_ : P\ S\ \times (DIM \times U) \rightarrow P\ S$$

$s' = s /<d',t'>=\{c/<d',t'>|c \in s\}$

**Definition 14** <u>*Choice.*</u> *This operator accepts two sets of contexts* $s1$, $s2$ *and non-deterministically returns one of them. The definition* $s= s1|s2$ *implies that s is either* $s1$ *or* $s2$.

$$\_|\_ : P\ S\ \times PS \rightarrow P\ S$$

**Definition 15** <u>*Override.*</u> *For every pair of context sets* $s1$, $s2$, $s1$, $s2 \in P\ S$ *this operator returns a set of contexts s, s* $\in P\ S$, *where every context* $c \in s$ *is computed as* $c1 \oplus c2$, $c1 \in s1$, $c2 \in s2$.

$$\_\oplus\_ : P\ S\ \times PS \rightarrow P\ S$$

$s' = s1 \oplus s2 = \{c1 \oplus c2 | c1 \in s1 \wedge c2 \in s2\}$

**Definition 16** <u>*Difference.*</u> *For every pair of context sets* $s1$, $s2$, $s1$, $s2 \in P\ S$ *this operator returns a set of contexts s, s* $\in P\ S$, *where every context* $c \in s$ *is computed as* $c1 \ominus c2$, $c1 \in s1$, $c2 \in s2$.

$$\_\ominus\_ : P\ S\ \times PS \rightarrow P\ S$$

$s = s1 \ominus s2 = \{c1 \ominus c2 | c1 \in s1, c2 \in s2\}$

Lifting the *undirected range* $\rightleftharpoons$ and *directed range* $\rightarrow$ to sets of contexts will produce higher-order sets. So, we do not define lifting for these two operators. However, since the results of applying these two operators are sets of contexts, the lifting operators can be applied to the results.

### 4.3.2 Relational Operators

We define the three relational operations $\bowtie$ (join), $\boxminus$ (intersection), and $\boxplus$ (union) for sets of contexts. In the following definitions, $c$ denotes a context, $si \in P\ S$ and $\Delta i = U_{c'} \in _{si} dim(c')$.

**Definition 17** <u>*Join.*</u>

$$\_\bowtie\_ : P\ S\ \times PS \rightarrow P\ S$$

$s = s1 \boxtimes s2 = \{c1 \cup c2 | c1 \in s1 \wedge c2 \in s2 \wedge c1 \downarrow \Delta 3 =$
$c2 \downarrow \Delta 3\}$, where $\Delta 3 = \Delta 1 \cap \Delta 2$.

**Definition 18** _Intersection._

$\_\boxdot\_ : P\ S \times P\ S \rightarrow P\ S$

$s = s1 \boxdot s2 = \{c1 \cap c2 | c1 \in c2 \wedge c2 \in s2\}.$

_We can prove that_ $s = s1 \boxdot s2 = (s1 \boxtimes s2) \downarrow \Delta 3$, _where_
$\Delta 3 = \Delta 1 \cap \Delta 2$.

**Definition 19** _Union._

$\_\boxplus\_ : P\ S \times P\ S \rightarrow P\ S$

$s = s1 \boxplus s2$ _is computed as follows:_

$\Delta_1 = \bigcup c \in s_1 dim(c),\ \ \Delta_2 = \bigcup c \in s_2 dim(c)$, _and_ $\Delta_3 = \Delta_1 \cap \Delta_2$

1. _Compute X1:_ $X1 = \{c_i \cup c_j \uparrow \Delta_3 / c_i \in s_1 \wedge c_j \in s_2\}$
2. _Compute X2:_ $X2 = \{c_j \cup c_i \uparrow \Delta_3 / c_i \in s_1 \wedge c_j \in s_2\}$
3. _The result is :_ $s = X1 \bigcup X2.$

Earlier we have shown that the results of $c_i = c_j$ and $c_i \rightarrow c_j$ are sets of contexts. So the relational operators $\boxtimes$ (join), $\boxdot$ (intersection), and $\boxplus$ (union) can also be applied to the expressions $c_i = c_j$ and $c_i \rightarrow c_j$, where $c_i$ and $c_j$ are contexts.

## 4.4 Context Set Expressions

Informally, a context set expression is an expression involving sets of contexts and context set operators. Let $s$ ranges over a set of contexts, $S$ over a context set expression and D over a dimension set. A formal syntax for context set expression $S$ is shown in Figure 2(left column).

| syntax | | precedence |
|---|---|---|
| $S ::= $ | $s$ $\mid$ $S$ | |
| $\mid$ | $S \oplus S$ $\quad$ $S \ominus S$ | $\downarrow, \uparrow, /$ |
| $\mid$ | $S \downarrow D$ $\quad$ $S \uparrow D$ | $\mid$ |
| $\mid$ | $S \boxtimes S$ $\quad$ $S \boxdot S$ | $\oplus, \ominus$ |
| $\mid$ | $S \boxplus S$ $\quad$ $S / \langle d, t \rangle$ | $\boxtimes, \boxdot, \boxplus$ |

■ Fig. 2 Formal Syntax of Context Set Expressions and Precedence Rules for Context Set Operators.

In order to precisely calculate a context set expression, we define the precedence rules for the context set operators. These are shown in Figure 2(right column) (from the highest precedence at the row to the lowest precedence in the bottom row). Parentheses will be used to override this precedence when needed. Operators having the same precedence will be applied from left to right.

## 4.5 Box Notation

In many applications it is of special interest to consider a set of contexts, all of which have the same dimension set and the tags corresponding to the dimensions in each context satisfy a given constraint. We use the notation _Box_ to denote such a set when the dimension set is $\Delta = \{d_1, ..., d_k\} \subset DIM$ and $p$ is a logical expression.

Note that in $p$, we are allowing the dimensions as variables, denoting the current tags. That is, if $p(d_1, d_2) = d_1 < d_2$, it means the current tag of $d_1$ is less than the current tag of $d_2$ in the context that has dimensions $d_1$ and $d_2$. A formal definition follows:

**Definition 20** _A Box set (or a Box for short) is a set of simple contexts with the same domain. Let_ $\phi \neq \{d_1, ..., d_k\} \subseteq DIM$ _be a set of dimensions and_ $p$ _be an expression in which the_ $d_i$ $(1 \leq i \leq k)$ _may occur as variables. Then_ $Box[d_1, ..., d_k / p] = \{c \in S / dim(c) = \{d_1, ..., d_k\}$ _and_ $p$ _is true when, for each I,_ $d_i$ _is assigned the value_ $c(d_i)\}$.

_The dimension_ $\Delta$ $(b)$ _of an onempty box_ $b$ _is the dimension of any (all) its elements._

The set of Box sets (or Boxes for short) are all sets of simple contexts all of which have the same domain. It is easy to show that anything defined by the Box notation is a Box.

## 4.6 Using Context Formalism in System Development

The two key terms in the study of context-aware systems are _context_ and _awareness_. Awareness is of two kinds. One kind is the internal monitoring of the system, called _self-awareness_ or _internal awareness_. System contexts are dynamic and consequently self-awareness varies from context to context. The other kind is the external monitoring of the system, called _external-awareness_. External awareness, also known as perception, is normally achieved through sensors and other stimuli, say from users or other system elements. External contexts change as and when the system environment changes, and such changes cause changes to external awareness. The system must use the knowledge it gained from its perception, apply it to the changing internals, and react by either triggering an internal state change or providing an external service. Hence, we must use a context formalism in which both self-awareness and externalawareness can be represented and reasoned about. Using context calculus we can compute dynamically different contexts, combine external and internal context, and calculate an internal context corresponding to an observed external context. Without the formalism such calculations can only be done in an ad hoc manner.

Context calculus has been implemented in C#. This context toolkit is portable and can be used as a plug-in for any context-aware application development. The component-based architecture given in [43] illustrates our approach to using context formalism for developing context-aware systems. Such an approach can be adapted to any context-aware application, including service-oriented systems [47], web services [48], and trustworthy systems [46].

**First A. Author** KaiYu Wan received her ph. degree of Computer Science from Concordia University in 2006, master degree in 1999 and bachelor degree in 1996 from NanJing University of Science and Technology. Currently she is a Lecturer at Department of Computer Science, East China Normal University. She has published 20 referred papers. Her research interests are in development of systems such as Trustworthy Systems, Component-Based Software Systems, Context-Aware System and Multi-Agent Systems, and in Programming Languages and Compilers. She enjoys reading and traveling.